# Conductive Heat Transfer through Nanoconfined Gas: From Continuum to Free-Molecular Regime

Reza Rabani [a*], Samy Merabia [b], Ahmadreza Pishevar [a]

[a] *Department of Mechanical Engineering, Isfahan University of Technology, Isfahan 84156-83111, Iran*
[b] *Univ Lyon, Univ Claude Bernard Lyon 1, CNRS, Institut Lumière Matière, F-69622, Villeurbanne, France*

**ABSTRACT:** In the past few decades, great efforts have been devoted to studying heat transfer on the nanoscale due to its importance in multiple technologies such as thermal control and sensing applications. Heat conduction through the nanoconfined gas medium differs from macroscopic predictions due to several reasons. The continuum assumption is broken down; the surface forces which extend deeper through the gas medium become prominent due to the large surface-to-volume ratio, and, finally, the gas molecules are accumulated nonuniformly on the solid surfaces. In this work, to better understand the combination of these phenomena on the heat conduction through the nanoconfined gas medium, we present a series of molecular dynamics simulations of argon gas confined between either metals or silicon walls. The gas density is set so that gas experiences a wide range of Knudsen numbers from continuum to the free molecular regime. It is observed that the intrinsic characteristics of the solid determine the gas density distribution near the walls and consequently in the bulk region, and these distributions control the heat conduction through the gas medium. While the nanochannel walls have their most significant impact on the density and temperature distributions of the rarefied gas, the pressure and the heat flux across the gas domain converge toward a plateau as the gas becomes denser. We propose new analytical formulas for calculating the gas pressure, induced heat flux, and effective thermal conductivity through the strongly nanoconfined gas, which incorporates the wall force field impacts on the gas transport characteristics for the Knudsen number in the range of 0.05 to 20.

[*] Email: reza.rabani@pd.iut.ac.ir (R.Rabani)

## A. Introduction

Growing demand for a better understanding of energy exchange on the nanoscale level has arisen during the past decade as nanotechnology is becoming an increasingly important part of contemporary sciences. Energy transfer between solid and gas medium in different rarefaction levels at the nanoscale attracts wide attention due to a broad range of scientific and engineering applications such as thermal diode [1–3], thermal switch [3–6], gas sensors [7–10], and energy harvesting applications [11–13]. Consequently, heat transfer across the interface of solid and gas has been studied continuously during past years using analytical, experimental, and numerical approaches [14–17]. Surface adsorption of the molecules on the solid walls, extension of the wall force field in the adjacent layer, and breakdown of the continuum description are the primary factors that complicate the numerical analysis of the nanoconfined gas medium. However, in order to tackle such phenomena at the atomistic level, recently, the molecular dynamics (MD) method is used extensively for studying heat transfer in nanoscales confined liquid [18–25] or gas medium [26–31].

The high surface-to-volume ratio of the confined fluid in nanochannels and nanoslit makes them the first candidate for mass and heat transfer in some special applications. Besides the above-mentioned intrinsic complexities, the high computational cost of MD simulations for rarefied gas in the nanochannel compared to its liquid counterpart is another limiting factor. This problem originates from the fact that the simulation domain should be extended at least one mean free path in the periodic directions. Hence, a larger simulation domain is required for the rarefied gas compared to the liquid ones [32]. The cold wall models such as Smart Wall Molecular Dynamics (SWMD) [32] and Virtual Wall Model (VWM) [33] have also been proposed to reduce computational costs for the MD simulation of confined gases. In cold wall models, wall atoms that do not interact with each other are frozen at their initial position, and only their interactions with the gas atoms are considered in the MD simulation. Therefore, the computational time decreases considerably.

A significant variation in the gas density and velocity was observed in the near-wall region when considering these cold wall models for the shear and force-driven nanochannel gas flow. Wall force field, the interaction force originates from the wall atoms which extends 1 nm through the gas medium, increasing the gas atoms' residence time in the near-wall region is responsible for this behavior [34,35]. Meanwhile, equilibrium simulations showed an anisotropic shear and normal stresses in the near-wall gas layer [36]. They also showed the role of the gas–wall interaction parameter and the Knudsen number (Kn), which is defined as the ratio of the mean free path ($\lambda$) to the characteristics nanochannel length scale ($l$), in the transport phenomena both in the bulk and near-wall region [37–39]. In addition, calculations showed that the gas–wall interaction energy controlled the surface adsorption of gas molecules [40]. Although many investigations well studied the hydrodynamics of confined gas in the nanochannels, less attention has been devoted to the thermal transport aspect of such configuration. In particular, the rarefied gas case is less explored due to the cold wall models' limitation and the high computational cost of real wall MD simulation.

Recently, heat conduction characteristics of the nanoconfined gas medium were studied thoroughly using the interactive thermal wall model (ITWM) [41]. Generally, the following phenomena were observed for heat conduction between two parallel walls:

1. Depending on the physical characteristics of the solid and gas atoms, the interfacial and wall interaction region thermal conductance could contribute significantly to the total thermal resistance of the gas, even for near micrometer-sized channels [42–44].
2. While the induced heat flux decreased as the walls become stiffer, increasing the solid-gas interaction energy enhanced it toward a plateau. In addition, an approximately bell-shaped type curve for induced heat flux was observed for heavy solid atoms [45].
3. The wall force field creates one peak in the distribution of the gas atoms at a distance of 1 nm from each wall while decreasing the wall temperature for the dense gas medium can extend this region and create a second peak [43].
4. The wall temperature can enhance or reduce the density increment in the wall interaction region for the cold and hot walls, respectively, by modifying gas residence time in the wall interaction region [46,47].
5. When the gas is dense, heat transport is purely diffusive. However, for rarefied gas, the heat transfer mechanism is a combination of ballistic transport in the bulk region and diffusive transport in the wall interaction region [43].

6. The number of adsorbed gas molecules on the wall is found to be a function of wall temperature and solid-gas interaction energy [43,44].

7. The gas medium's kinetic stress, particle-particle virial, and surface-particle virial are affected by the gas density, wall temperature, and induced heat flux. As the gas becomes hotter, the normal stress distribution increases accordingly due to the increment in the kinetic stress, which dominates the bulk region. Near the hot wall, the surface-particle and particle-particle virial stress are attenuated, while near the cold wall, these virials are intensified [48].

8. Due to the anisotropic distribution of the normal stress components in the wall interaction region of nanoscale-confined gas, it is impossible to define a pressure in such region. However, the normal stress distribution is uniform in the bulk gas region which allows the definition of the gas pressure there [36,48].

9. Thermal conductivity is found to be a local property that depends on the distance from the wall. The highest value of the local thermal conductivity is observed in the bulk region of the gas domain while the lowest was found in the wall interaction region [42,43,47]. In order to be able to compare the overall thermal conductivity of the gas domain, an effective thermal conductivity concept is used [42,49] in which the effect of all gas layers is included.

These observations were obtained based on a series of parametric studies in which one parameter was changed while the other parameters were kept constant. However, for the real wall in nanochannels, the dominating parameters such as wall stiffness, wall atom mass, and solid-gas interaction energy change simultaneously based on the wall material. Therefore, the phenomena mentioned above might change accordingly. In this study, to have a better understanding of the heat transfer characteristics across the nanoconfined gas domain, Silver (Ag), Aluminum (Al), Gold (Au), Copper (Cu), Platinum (Pt), and Silicon (Si) are considered as wall materials and argon considered as the gas medium. The gas density has been changed while the channel height is kept constant at 5.4 nm, so we could simulate a wide range of gas rarefaction levels from the continuum to free molecular regimes. This study enables us to estimate the combined effect of metal and nonmetal wall material and the gas density on the heat transfer characteristics of the nanoconfined gas domain.

## B. Computational Model

Simulations are performed using the open-source software LAMMPS (Large-Scale Atomic/Molecular Massively Parallel Simulator) [50]. A schematic sketch of the simulation domain for two different levels of gas rarefaction is shown in Fig. 1. In the axial (x) and lateral (z) directions, periodic boundary conditions are applied. Meanwhile, 350 K and 300 K are assigned to the bottom (hot) and top (cold) walls, respectively. In order to make sure that our data is independent of the simulation domain size, it is extended at least for one mean free path in (x) and (z) directions [32]. To model the van der Waals interactions between gas-gas and gas-wall atoms, the truncated (6–12) Lennard–Jones (L–J) potential is used

$$\phi(r_C) = 4\varepsilon \left( \left( \left(\frac{\sigma}{r_{ij}}\right)^{12} - \left(\frac{\sigma}{r_{ij}}\right)^{6} \right) - \left( \left(\frac{\sigma}{r_c}\right)^{12} - \left(\frac{\sigma}{r_c}\right)^{6} \right) \right), \tag{1}$$

where $r_c$ is the cutoff distance, $r_{ij}$ is the interatomic distance, and $\phi(r_C)$ is the value of the interatomic potential at r=r$_C$. The molecular diameter is $\sigma = 0.3405$ nm, depth of the potential well for argon is $\varepsilon = 1.65 \times 10^{-21}$ J, and the mass of an argon atom is $m = 6.63 \times 10^{-26}$ kg. Because the L-J potential is negligible at a large molecular distance, the cut-off radius is commonly set to r$_C$ = 2.7 nm and 1.08 nm for the dense and dilute gas medium, respectively [36]. The interaction parameters between wall and gas atoms are listed in Table 1, which is obtained based on the Lorentz-Berthelot (L-B) mixing rules for interaction strength, i.e., $\varepsilon_{solid-gas} = \sqrt{\varepsilon_{solid-solid} \times \varepsilon_{gas-gas}}$. The embedded atom method (EAM) was used to model the metallic wall-wall interactions, which is a many-body interaction potential and works very well for FCC metallic structures [51]. Calculating the embedding energy as a function of the atomic electron density enables this method to predict a metallic structure's total energy accurately. Meanwhile, the Si-Si interactions were calculated using the Stillinger–Webber (SW) potential, which applies two-body terms for the pairwise interactions with additional many-body terms to stabilize the structure [52].

Newton's second law determined the motion of the gas and wall atoms by applying a velocity Verlet time integration algorithm [53]. The NVE ensemble is adopted in all simulations, and the LANGEVIN thermostat sets the desired temperature at the wall. Therefore, the thermostat is not applied to the gas when heat is appropriately transferred through the walls. The Maxwell–Boltzmann velocity distribution with the mean temperature of 325 K is used to initialize the argon gas and wall atoms at the initial state of the simulations. While a timestep of 4 fs is used in all simulations, smaller timesteps were also tested and had no significant effect on the results. The thermal equilibrium is reached when the initial particle distribution evolved for $n_{init}$ timesteps. Then, the temperature is set to 350 K and 300 K for the bottom and the top wall, respectively, and the system is allowed to relax towards the expected temperature in $n_{ssc}$ timesteps. Finally, additional $n_{ave}$ timesteps are performed before the microscopic quantities are averaged to obtain the macroscopic properties.

The value of $n_{init}$, $n_{ssc}$, and $n_{ave}$ for the rarefied gas medium are fixed to 5, 10, and 20 x $10^6$ timesteps, respectively, while for the dense gas medium 5, 10, and 20 x $10^5$ timesteps are performed. Longer time averaging is also performed in all cases to confirm the convergence of calculated quantities. Notable to say that the reported macroscopic observations in this study are the average over four and eight independent simulations for dense and rarefied gas medium, respectively, where the random generator number for the initial gas atoms distribution has been changed. The channel height is determined from the centerlines of the first layer of wall atoms at the top and bottom surfaces. The computational domain is divided into bins of approximately 0.054 nm and 0.108 nm for the density and temperature, respectively, perpendicular to the walls (y) direction.

We calculate the gas pressure in the middle of the channel as [36]

$$P = (S_{xx} + S_{yy} + S_{zz})/3, \qquad (2)$$

where $S_{xx}$, $S_{yy}$ and $S_{zz}$ are the normal stress components in $x$, $y$, and $z$ directions, respectively, and are calculated as:

$$S_{kl} = \frac{1}{Vol}\left(\sum_{i=1} m_i(V_i^\alpha - \overline{V}_i^\alpha)\left(V_i^\beta - \overline{V}_i^\beta\right) + \frac{1}{2}\left(\sum_{i,j} r_{ij}^\alpha\right)f_{ij}^\beta\right), \qquad (3)$$

where $\alpha$ and $\beta$ are the axes of the Cartesian coordinate system. In addition, $\overline{V}_i^\alpha$ and $\overline{V}_i^\beta$ are the local average streaming velocities at the location of particle $i$ in $\alpha$ and $\beta$ directions, respectively, and $V_i^\alpha$ and $V_i^\beta$ are peculiar velocity components in $k$ and $l$ directions. While $f_{ij}^\beta$ is the $\beta^{\text{th}}$ component of the force exerted on the atom $i$ from the atom $j$, $r_{ij}^\alpha$ is $\alpha^{\text{th}}$ component of the relative distance between particles $i$ and $j$.

The Irving–Kirkwood (I–K) expression is applied to determine the induced heat flux vector through the gas domain [54,55]:

$$J_l = \frac{1}{\text{Vol}} \langle \sum_i V_l^i E_{tot}^i + \frac{1}{2} \sum_{i,j} r_l^{ij}(f^{ij}.V^i) \rangle, \qquad (4)$$

$$E_{tot}^i = \frac{1}{2} m^i \left( (V_x^i)^2 + (V_y^i)^2 + (V_z^i)^2 \right) + \emptyset^i, \qquad (5)$$

where the summation is performed over all gas atoms. Considering $\beta$ as the axes of the cartesian coordinate system, $V_l^i$ is the velocity component of particle $i$ in the $\beta$-direction. In addition, $E_{tot}^i$ is the total energy, and $\emptyset^i$ is the potential energy of particle $i$, which is calculated using equations (5) and (1), respectively, and $r_l^{ij}$ is the distance vector between particle $i$ and $j$. Furthermore, $f^{ij}$ is the vector of intermolecular force applied to particle $i$ by particle $j$, and $V^i$ is the velocity vector.

### C. Result and Discussion

The simulations are performed for two different sets of physical situations. In the first set of simulations, we changed the number of the argon gas atoms in the simulation domain to obtain the same Knudsen for the gas in the middle of the channel for all different walls. The number of gas atoms adsorbed on the walls is mainly affected by the solid-gas interaction energy, which is not the same for all wall materials. Therefore, the exact number of gas atoms leading to the desired density in the middle of the channel is found via a trial-and-error method for each wall material. For the rarefied gas medium, the Knudsen number in the middle of the channel is set at 11 while for the dense gas case, 0.11 is considered. It should be mentioned that the Knudsen number is defined based on the gas density in the middle of the channel. From here on, we call this set of simulations the constant Knudsen case. On the other hand, in the second set of simulations, the number of gas atoms in the simulation domain is kept constant which is called the constant atom number case. For this set of simulations, 10000, 30000, 60000, and 100000 gas atoms are considered in the simulation domain, which leads to distinct Knudsen numbers in the middle of the channel and consequently different heat conduction characteristics.

Figure 2 shows the gas density distribution across the channel height for different walls of the constant Knudsen case. It is interesting to notice that for the rarefied gas regime in Fig. 2a, a nanochannel with Pt walls requires 27825 gas atoms to reach the Knudsen number of 11 in the middle of the channel while for the Si walls, this reduces to 1000 gas atoms only. Other metals require many gas atoms between these limiting conditions. The difference originates from the different gas/metal interaction strengths, as shown in Table 1. Higher interaction strength between gas/metal compared to gas/gas atoms leads to a denser gas layer in the wall interaction region, i.e., 1 nm from each wall. As an example, the ratio between maximum gas density in the wall interaction region near the cold wall and the gas density in the middle of the channel is approximately 700 and 15 for the Pt and Si channel, which shows that the selected materials cover a wide range of physical phenomena.

Figure 2 shows that for all the metallic walls, the maximum gas density occurs at a distance of 0.5 nm from the solid wall surface. In contrast to the Si wall, the maximum gas density formed closer to the surface. It means that the gas atoms get closer to the Si walls as compared with metallic ones. A closer look at the wall's physical characteristics should reveal the source of such differences. Table 2 shows the physical characteristics of the different materials. The solid stiffness, $K_S$, corresponding to the spring constant, is obtained for each material as follows [56,57]:

$$K_S = \frac{m_w K_B^2 T_E^2}{\bar{h}^2}, \tag{6}$$

$T_E$ is the Einstein temperature, $m_w$ is the solid atom mass, $K_B = 1.3806 \times 10^{-23} \, J/K$ is the Boltzmann constant, and $\bar{h}$ is the reduced Planck constant. Notably, the Einstein temperature is derived based on the Debye temperature of the wall materials at room temperature of $298 \, K$ [58,59]. Table 2 clearly shows that the Si wall has the highest stiffness compared to the metals. For the stiffer wall, the atoms oscillate with lower amplitudes (~1/$K_S$) around their equilibrium position [60,61] letting the gas atoms come closer to the wall as shown in Fig. 2a for the silicon wall.

According to Fig. 2b, the same phenomena are observed for the dense gas condition of the constant Knudsen case. The Knudsen number of 0.11 in the middle of the Pt and Si channel is achieved by the most 89000 and the least 67000 number of gas atoms, respectively. While Cu and Pt have the highest gas densities in the cold wall interaction region around 2000 kg/m$^3$, Si has the lowest around 1250 kg/m$^3$ near the cold wall. On the other hand, the ratio between maximum gas density in the wall interaction region of the cold wall and the middle of the channel is approximately 9 and 6 for Pt and Si walls. This ratio decreased too much compared to the rarefied gas, as discussed above. This comparison reveals that while the wall force field still affects the distribution of the atoms in the dense gas medium, its effects are much more pronounced for the rarefied condition. The reason behind such a phenomenon is that for very low-density gas conditions, a gas atom mainly interacts with the force fields of the wall atoms rather than neighboring gas atoms. Therefore, the wall force field almost controls the motion of gas atoms. As the gas becomes denser, the number of the neighboring gas atoms increased accordingly, and consequently, the contribution of the gas atom's force field is enhanced. Hence, in such a case, both the solid wall and the neighboring gas atom's force field are the parameters that determine the motion of gas atoms. Figures 3a and 3b show the temperature distribution across the channel height for the rarefied and dense gas cases, respectively.

Figure 3a clearly shows that the temperature profile of dilute gas for metallic walls may be decomposed in three distinct regions. These divisions are represented exclusively in Fig. 4a. In the first region, called the adsorbed gas layer, which extended from 0.2 to 0.7 nm from the wall the gas and the wall atoms are in thermal equilibrium with the same temperature. The second region extends from 0.7 to 1.5 nm near each wall and is called the wall interaction region. In this layer, a linear variation of the temperature profile is observed. Beyond 1.5 nm from the wall, the third layer is observed as a bulk region, i.e., the region in the middle of the channel where the gas atoms do not feel the wall force field is almost at a constant temperature. The distinct behaviors originate from the different heat transfer mechanisms of the gas domain in these regions [66]. In the wall interaction region, the Knudsen number varies between 0.08~0.016 (corresponding to the variation of gas density varies from 300 to 1500 kg/m$^3$), depending on the metal type and temperature, which is responsible for the diffusive heat transfer mechanism by gas atoms collision. In the bulk gas region, the gas Knudsen number is around 11 for all the metal walls, leading to a ballistic heat transfer mechanism as collisions between the gas atoms are rare in the middle of the channel. It is clear in Fig. 3b that unlike the rarefied case, the temperature profile of dense gas shows two distinct regions for metallic walls. The first region is the same as the rarefied condition, i.e., an adsorbed gas layer, which extends from 0.2 to 0.7 nm, and the gas has the same temperature as the walls. The rest of the channel height can be considered the second region. A linear variation of the temperature between the top and bottom walls is a signature of the diffusive heat transfer mechanism which dominates in this region.

Inspection of Figs. 3a and 3b reveal that for the Si channel, the gas temperature profile consists of two distinct regions for the rarefied gas and only one region for the dense gas. As shown in Fig 3a for the temperature profile of the rarefied gas between silicon walls, the adsorbed gas layer region is disappeared while the wall interaction region extends from 0.2 to 1 nm where linear variation for the temperature profile is observed. The rest of the channel height is dominated by the bulk medium for which the uniform temperature is a signature of the presence of ballistic heat conduction mechanism in this region. Meanwhile, for the dense gas case in Fig 3b, a temperature jump of 4 K occurs between the gas and Si surface, and a linear temperature profile is also observed between the gas temperature adjacent to the hot and cold walls. In contrast to metallic walls for which the temperature distribution along the channel height can be expressed with a unique curve regardless of the material type, see Figs. 3a and 3b, the Si wall temperature profile behaves quite differently. Figs. 2a and 2b imply that besides the gas density distribution in the bulk region, the distribution of the gas atoms in the wall interaction region is the other critical parameter that characterizes the temperature distribution across the channel height.

Figure 5 shows the density profile along the channel height for the constant gas atom number case. Since the number of the gas atoms is kept constant, the gas density in the middle of the channel is not the same for the selected wall materials due to the different solid-gas interaction energy. It can be observed that from all four cases, the gas in the Pt and Cu nanochannels produce the highest gas density near each wall while the Si has the lowest density. Our observation is a direct consequence of stronger interaction between Pt/Cu and the gas atoms in comparison with Si as stated in Table 1. Since the total number of the gas atoms is constant in each simulation, as expected the lowest gas density in the middle of the channel belongs to the Pt and Cu nanochannels and the highest one belongs to the Si nanochannel. Considering the gas density in the bulk region of the channel for the most dilute gas case in Fig. 5a, the gas Knudsen number in the middle of the channel for the Pt nanochannel is approximately 434 while it is 1.06 as it comes to the Si walls which shows orders of magnitude differences. On the other hand, for the densest gas medium in Fig. 5d, the bulk gas Knudsen number for the Pt nanochannel is approximately 0.089 and changes to 0.067 for the Si ones which shows the same order of magnitude. This analogy reveals that the effect of the wall material on the density distribution of the rarefied gas is much more pronounced than the dense ones, as observed earlier for the constant gas density case.

It is interesting to notice in Fig. 5a that for the most dilute gas case and the metallic wall, the maximum gas density near the cold and hot walls is in the range of 600~850 kg/m$^3$ and 275~350 kg/m$^3$, respectively, depending on the metal type. However, for the densest gas medium of Fig. 5d, the maximum gas density near the cold and hot walls changes in the range of 1500~2000 kg/m$^3$ and 1400~1900 kg/m$^3$. The same trend is also observed for the Si nanochannel. The results notify that for the most dilute case, the gas near the cold wall is twice denser than that of the hot wall, while for the highest density case, it is somehow in the same order of magnitude. Therefore, it is deduced that the impact of wall temperature on the dilute gas is more significant than on the dense gas.

Figure 6 shows the temperature distribution in gas along the channel height for the constant number of gas atom cases. A gradual transition from a uniform temperature to a linear temperature profile can be observed in the bulk region as the Knudsen number moves toward the continuum regime, moving from Fig. 6a to Fig. 6d. Meanwhile, an identical temperature profile for the different metallic walls is formed while the Si wall has a different temperature distribution regardless of the gas Knudsen number. Moreover, a temperature jump occurs even near the continuum regime in Fig. 6d.

Shi et al. [84] showed that in the nanoconfined liquid medium, the temperature could not be defined at some location in the liquid argon adjacent to the solid surface. However, we have shown here that for a wide range of gas density and physical characteristics of the wall, the temperature can be defined in any position across the channel height, as shown in Fig. 6. This difference refers to the single peak nature of the density distribution in the nanoconfined gas compared to the density layering phenomena in the liquid medium.

The gas pressure in the middle of the channel is calculated using equations (2) and (3) as depicted in Fig. 7. The bulk pressure is the same for the various wall material in the constant Knudsen case, as shown in Fig. 7a. For the rarefied gas case, the bulk pressure is about 150 kPa while for the dense gas medium, it is about 14 MPa. Figure 7b shows the calculated pressure for the constant atom number case. For 10000 gas atoms, the Pt and Si walls have the lowest and highest gas pressure in the middle of the channel of 3.77 kPa and 1.54 MPa, respectively, which shows a notable difference. The pressure developed by the other metals is between these limiting ranges. However, by increasing the number of gas atoms to 100000, the lowest and highest values of pressure by the Pt and Si walls in the middle of the channel are in the same order of magnitude and are 18 to 24 MPa, respectively. The other metals lead to the pressure of around 20 MPa in the gas. This observation again proves that the wall force field has its highest effect on the rarefied gas condition.

Using equations (4) and (5), the induced heat flux in the simulation domain is also computed for rarefied and dense gas cases. Fig. 8a shows that for the rarefied gas in the constant gas Knudsen number case, the highest and lowest induced heat flux is observed for Al and Si walls, 4 and 3.36 MW/m$^2$ respectively, while for the other metallic walls is around 3.85 MW/m$^2$. Similarly, for the dense gas, the maximum and minimum induced heat flux again belong to the Al and Si walls which are in the order of 178 and 150 MW/m$^2$ respectively, while it is 174 MW/m$^2$ for the other metals. We deduced from the results that as long as the gas Knudsen number is kept constant in the middle of the channel, the heat flux is not significantly affected by the material type for all metallic walls. However, the estimation is lower for the Si wall. The slight difference in the calculated heat flux for the metallic walls can be assigned to the inexact implementation of the thermostat applied to the walls and its adjacent fluid layer [85–87]. In the current study, the wall temperature is monitored carefully during the simulations for all cases to see whether the thermostat produces the desired temperature on the walls. The results show that the wall temperature is set with a maximum deviation of 1 K to its predefined values.

The maximum deviation in the heat flux for the nanochannels with metallic wall considered here is for the Al for which the walls and its adjacent gas layer experiences a temperature of 301 and 351 K, as shown in Fig. 3. The observed deviation in the wall temperature leads to a slightly higher heat flux for the rarefied and the dense gas cases, as shown in Fig. 8a. However, the difference of Si wall heat flux with that of the metallic walls is so remarkable that it cannot be related to the applied thermostat. Actually, for the Si walls, the maximum gas density near each wall is much lower than its corresponding metallic walls, according to Fig. 2. Therefore, the number of collisions between the gas and the wall atoms decreased, reducing the transferred heat flux and resulting in a temperature jump, as shown in Fig. 3.

As stated before, the solid/ gas atom mass ratio, solid/gas interaction energy, and solid stiffness are the key parameters that affect the gas density distribution near the surface and eventually modify energy transfer at the solid/gas interface [67,68,82,83]. The values used for these parameters in our simulations are listed in Table 2. Interestingly, despite parameter differences, the heat flux through the rarefied and dense gas is approximately the same for all metallic walls, as shown in Fig. 8a. The accumulation of the gas atoms in the wall force field can be the origin of this observation. The number of gas atoms adjacent to the wall is increased so that the number of collisions between gas and solid atoms becomes independent of the parameters mentioned above. Therefore, the adsorbed gas on the wall reaches thermal equilibrium with the wall, as shown in Fig. 3.

Figure 8b shows the heat flux for the constant number of gas atom cases. Here, for the dilute gases, the extreme cases correspond to Pt walls with a heat flux of 0.112 MW/m$^2$, and the Si walls with 28.6 MW/m$^2$. Such extraordinary difference originates from the different gas Knudsen numbers in the middle of the channel, as shown in Fig. 5a. As the number of gas atoms increases in the densest case, Pt and Al correspond to the lowest and highest heat flux of 203 and 232 MW/m$^2$ through the gas domain. However, such huge differences disappeared, and the induced heat flux for all wall materials converged toward a plateau between these limiting conditions.

Our discussion of induced heat flux is based on the converged value of gas density in the middle and near the channel walls so far. However, it is of practical use to reshape the obtained results for the heat flux in terms of the gas Knudsen number in the middle of the channel. Therefore, the calculated heat flux for both sets of simulations is rearranged in Fig. 9a based on their corresponding Knudsen number. Such representation helps us introduce a new correlation for the heat flux in the nanochannels for a wide range of Knudsen numbers between 0.05 to 20, which covers the whole transition regime in addition to the continuum and free molecular regimes. Equation (7) represents an analytical formula for the calculation of the induced heat flux through the gas domain in the transition regime between ballistic and diffusive regimes, with thermal conductivity of K, between two parallel surfaces at the temperature of $T_1$ and $T_2$ in the form of [62]:

$$q'' = \frac{K(T_1 - T_2)}{l\left(1 + \frac{2-\alpha}{\alpha}\frac{9\gamma-5}{\gamma+1}\sqrt{\frac{T_{m,FM}}{T_{m,DF}}} Kn\right)}, \tag{7}$$

in which $\alpha$ denotes the energy accommodation coefficient, $l$ is the distance between the surfaces, $\gamma$ is the gas heat capacity ratio defined as $C_P/C_v$, and $T_{m,DF} \sim T_{m,FM} \sim (T_1 + T_2)/2$ when the temperature difference between the surfaces is smaller than the temperature of the cooler ones, as we have in our study. This equation is derived analytically in which the effect of the wall force field in the density distribution and its corresponding impact on the induced heat flux is absent. By fitting this equation into our current data, we can incorporate the effect of the wall force field into the heat flux correlation for the nanochannels. For the argon as working media, $\gamma$ equals 1.667 at 300 K, and the thermal conductivity $K$ is estimated in terms of gas density by the semi-empirical equation of Lemmon and Jacobsen $0.027\rho + 17.8$ [63].

Experimental investigation shows that $\alpha$ for the argon gas interacting with metallic and silicon walls is about 0.9~0.95 for the ordinary surface without any special treatment [64]. It was also shown that for a cleaner surface with argon-plasma treatment, these values are decreased by 0.05 [64]. More reduction in $\alpha$ might occur for a much cleaner surface, which is hard to achieve. Actually, in practical application, surface roughness and contamination, such as adsorbed gas atoms on the solid surface, enhance the value of $\alpha$ toward one [65]. It is also shown that using the L-B mixing rule to calculate the solid-gas interaction energy, as used in the current study, usually leads to $\alpha \sim 1$ which is a good approximation for the practical application. Therefore, assuming $\alpha = 1$ for the solid-gas interactions, equation (7) can be rewritten as follows:

$$q'' = \frac{0.27(\chi\rho) + 178}{1 + 3.75(\beta Kn)}, \tag{8}$$

where $\chi$ and $\beta$ are dimensionless parameters. Substituting the gas density by $\rho = 23.85/Kn$ in equation (8), as proposed for the argon gas [43], the heat flux can be expressed in terms of gas Knudsen number as:

$$q''(MW/m^2) = \frac{178 + \frac{8.85}{Kn}}{1 + 4.54 Kn}, \tag{9}$$

For parameters $\chi = 1.375$ and $\beta = 1.21$, equation (9) is compared to numerical data in Fig. 9a.

The calculated heat flux is also used to calculate the effective thermal conductivity concept [42,49] for the gas domain as shown in Fig. 9b. Clearly, the overall trend is the same as the induced heat flux curve. This quantity can be expressed as the following equation:

$$K_{eff}(mW/m-K) = \frac{19.22 + \frac{0.956}{Kn}}{1 + 0.49 Kn}, \tag{10}$$

Based on the effective thermal conductivity, the induced heat flux for other channel widths and the temperature difference can be also calculated using Fourier's Law provided that the wall interaction region covers a significant portion of channel height (here it is 40%).

The calculated gas pressures in the middle of the channel are also rearranged based on their corresponding Knudsen number as shown in Fig. 10. Considering the similarity in the distribution of the induced heat flux (Fig. 9a) and the gas pressure (Fig. 10), an analytical formula similar to equation (9) is also obtained for the gas pressure in the middle of the channel as follows:

$$P(MPa) = \frac{0.23 + \frac{1.63}{Kn}}{1 + 0.16 Kn}, \tag{11}$$

Equations (9)-(11) are the universal curve for calculating the heat flux, effective thermal conductivity, and gas pressure for the argon gas which can be used in a wide range of Knudsen numbers from continuum to rarefied gas conditions.

Finally, the compressibility of the nanoconfined gas is shown in Fig. 11 defined by $Z=P/\rho RT$ in which P and ρ are taken as the pressure and density of the gas in the middle of the channel and R is the gas constant for the argon equals 0.208 kJ/kg-K. It is interesting to notice that for all cases 0.98<Z<1.02 which means that the gas behaves as an ideal gas in the bulk region of the nanochannel. However, it should be noted that due to the anisotropic distribution of the normal stress in the wall interaction region, argon behaves as a non-ideal gas medium in the near-wall region [36,48].

## Conclusions

Heat conduction through the nanoconfined argon gas medium has been studied thoroughly using molecular dynamics simulation. Ag, Al, Au, Cu, Pt, and, Si are considered the wall materials in a 5.4 nm channel height which means that the wall interaction region covers 40% of the channel height. The gas density has been changed while the channel height is kept constant, so a wide range of Knudsen numbers from continuum to free molecular regime has been spanned. The main objective of this study is to provide a better insight into how the interplay between the physical characteristics of the wall and the wall interaction region affects the thermal behavior of strongly nanoconfined gas. The following phenomena are observed:

- The gas density distribution along the channel height is mainly affected by the physical characteristics of the wall when the gas is rarefied. Generally, the density peak near the silicon wall is weaker than the metallic ones, forming closer to the wall.
- The temperature profile for the rarefied gas in the metallic nanochannel consists of an adsorbed gas layer, which is in thermal equilibrium with the adjacent wall, a wall interaction region in which the diffusive transport mechanism is dominated, and a bulk region with a ballistic transport property. For the dense gas case, the adsorbed gas layer and bulk region with diffusive nature are only observed. For the silicon wall, the general trend is the same while the adsorbed gas layer vanishes.
- The effect of wall material on the pressure, heat flux, and effective thermal conductivity of the gas domain are much more pronounced as the gas is rarefied. These parameters are described by master curves that can be used for a wide range of Knudsen numbers from continuum to free molecular regime. The master curves converge toward a plateau as the gas becomes denser which shows that the wall material type loses importance.

- While gas shows non-ideal behavior in the wall interaction region, it is shown that strongly confined argon at 325 K behaves as an ideal gas in the bulk region of the channel for the Knudsen number in the range of 0.05 to 20.

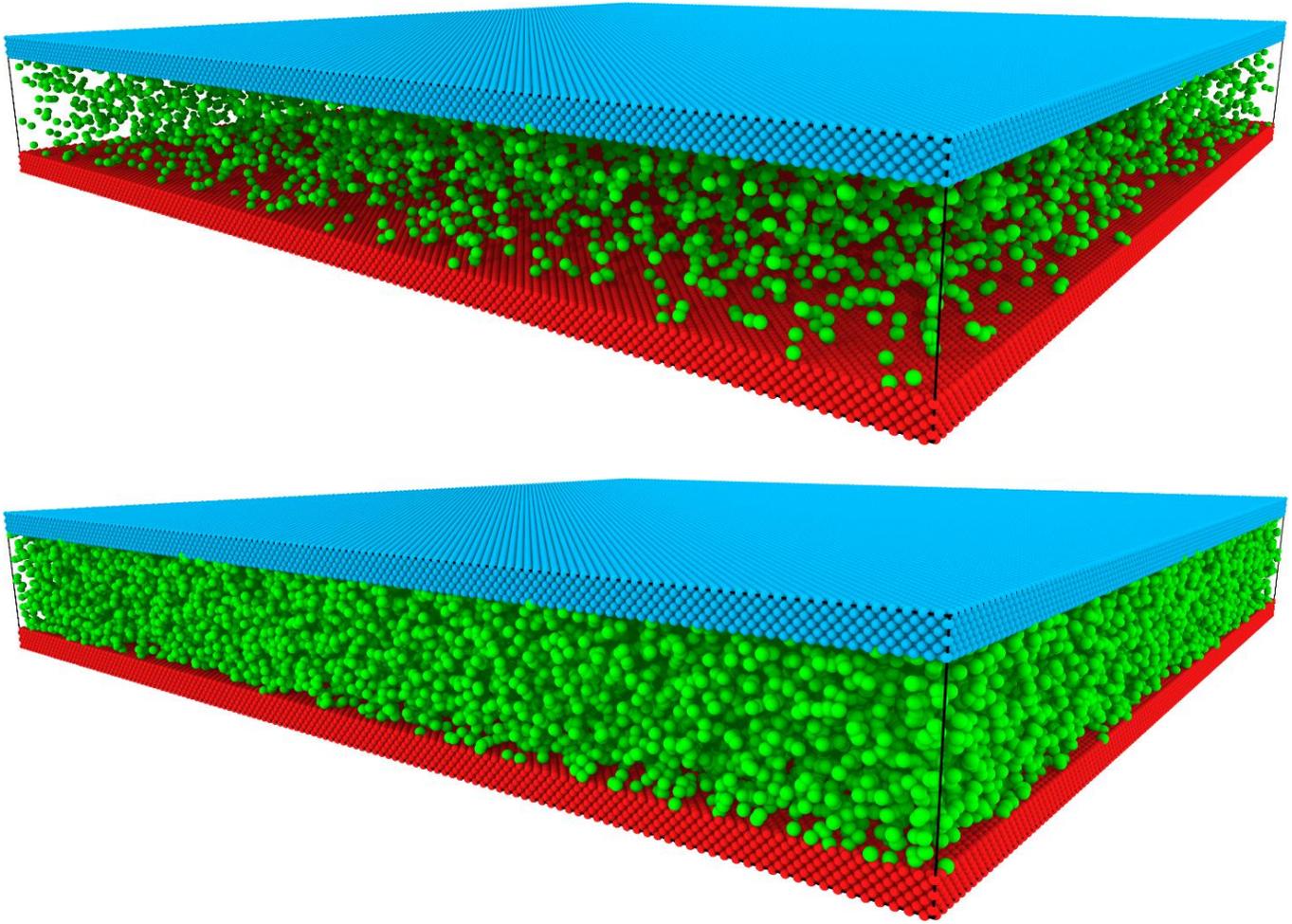

**Fig. 1.** Snapshot of the simulation domain for dilute **(top)** and dense **(bottom)** gas medium between top $(Cold - 300\ K)$ and bottom $(Hot - 350\ K)$ walls

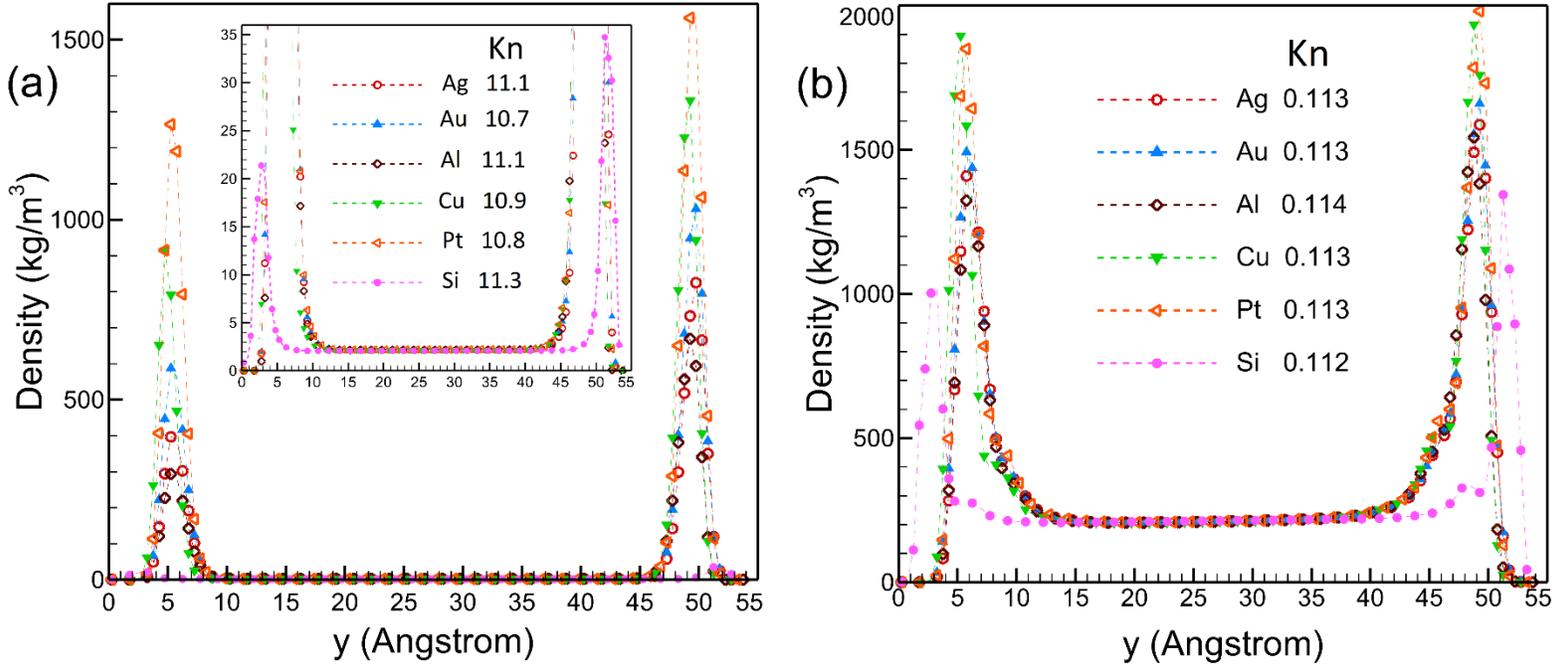

**Fig. 2.** Density distribution in the simulation domain for dilute (Ag 13100 - Au 17000 – Al 10500 – Cu 20200 – Pt 27825 – Si 1000) **(a)** and dense (Ag 71000 - Au 83000 – Al 78500 – Cu 284000 – Pt 89000 – Si 67000) **(b)** gas medium for constant Knudsen case (Knudsen numbers are computed based on the gas density in the middle of the channel)

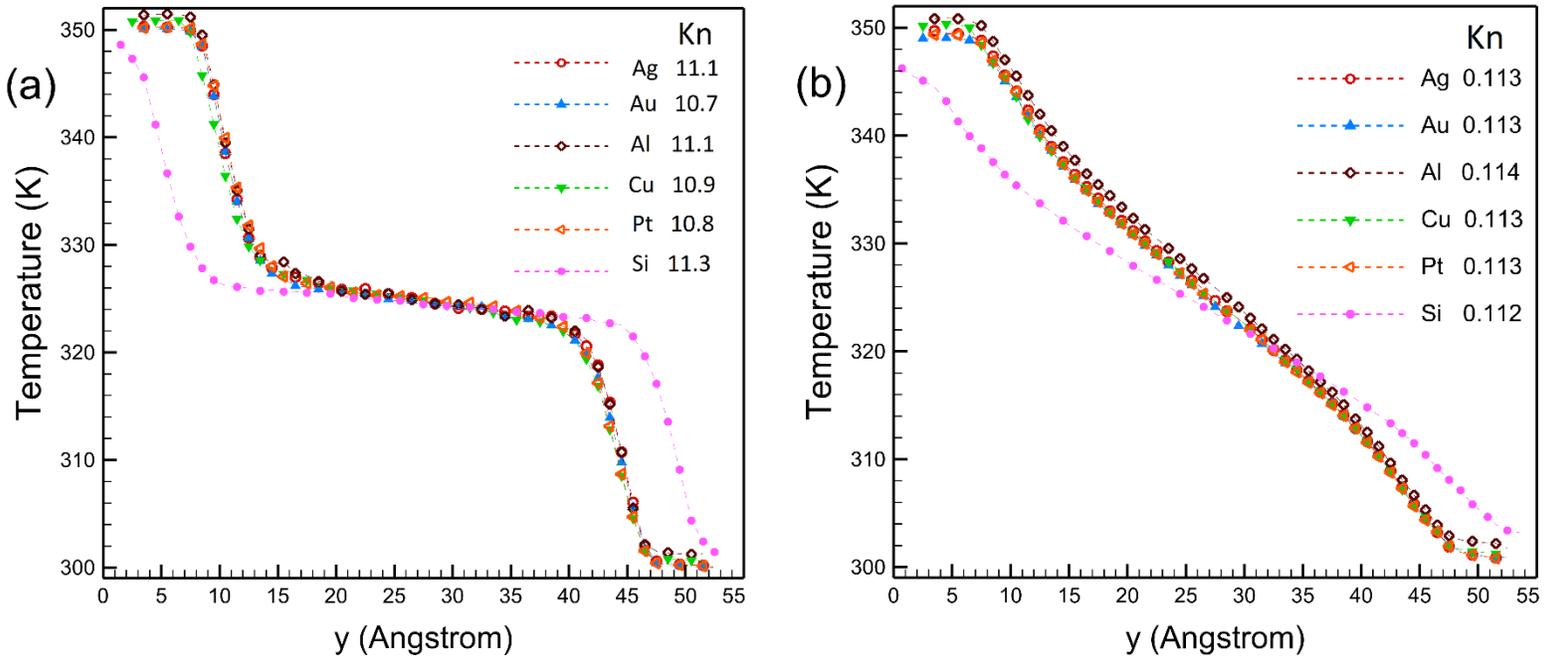

**Fig. 3.** Temperature distribution in the simulation domain for dilute (Ag 13100 - Au 17000 – Al 10500 – Cu 20200 – Pt 27825 – Si 1000) **(a)** and dense (Ag 71000 - Au 83000 – Al 78500 – Cu 284000 – Pt 89000 – Si 67000) **(b)** gas medium for constant Knudsen case (Knudsen numbers are computed based on the gas density in the middle of the channel)

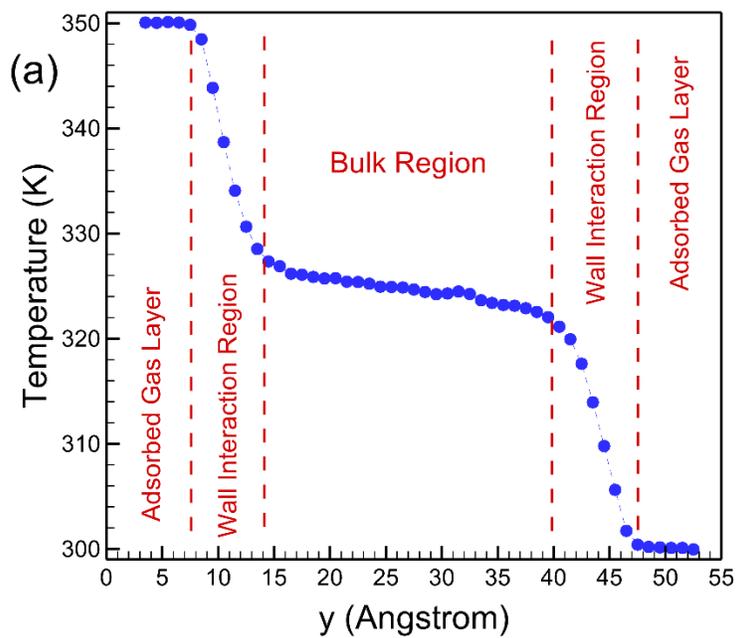 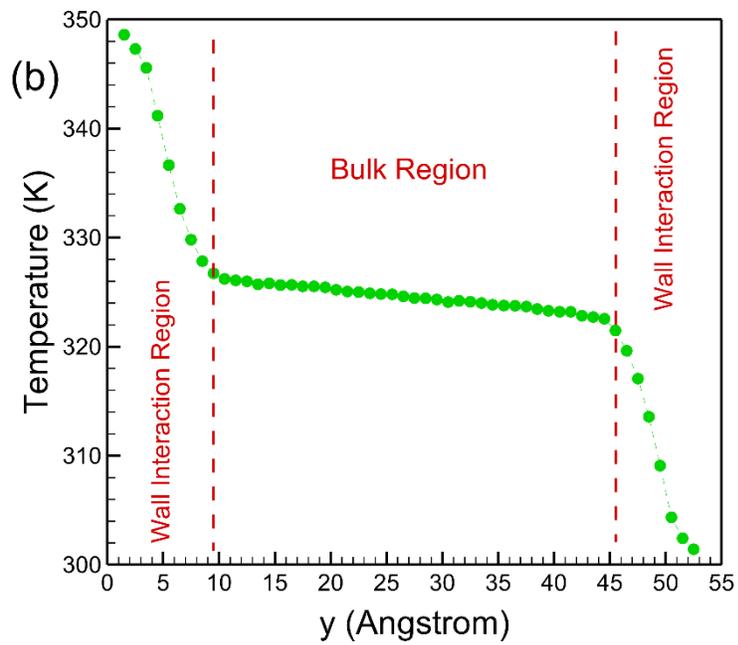

**Fig. 4.** Schematic division of the channel height based on the transport mechanism for metallic **(a)** and Silicon **(b)** walls based on the different transport regime

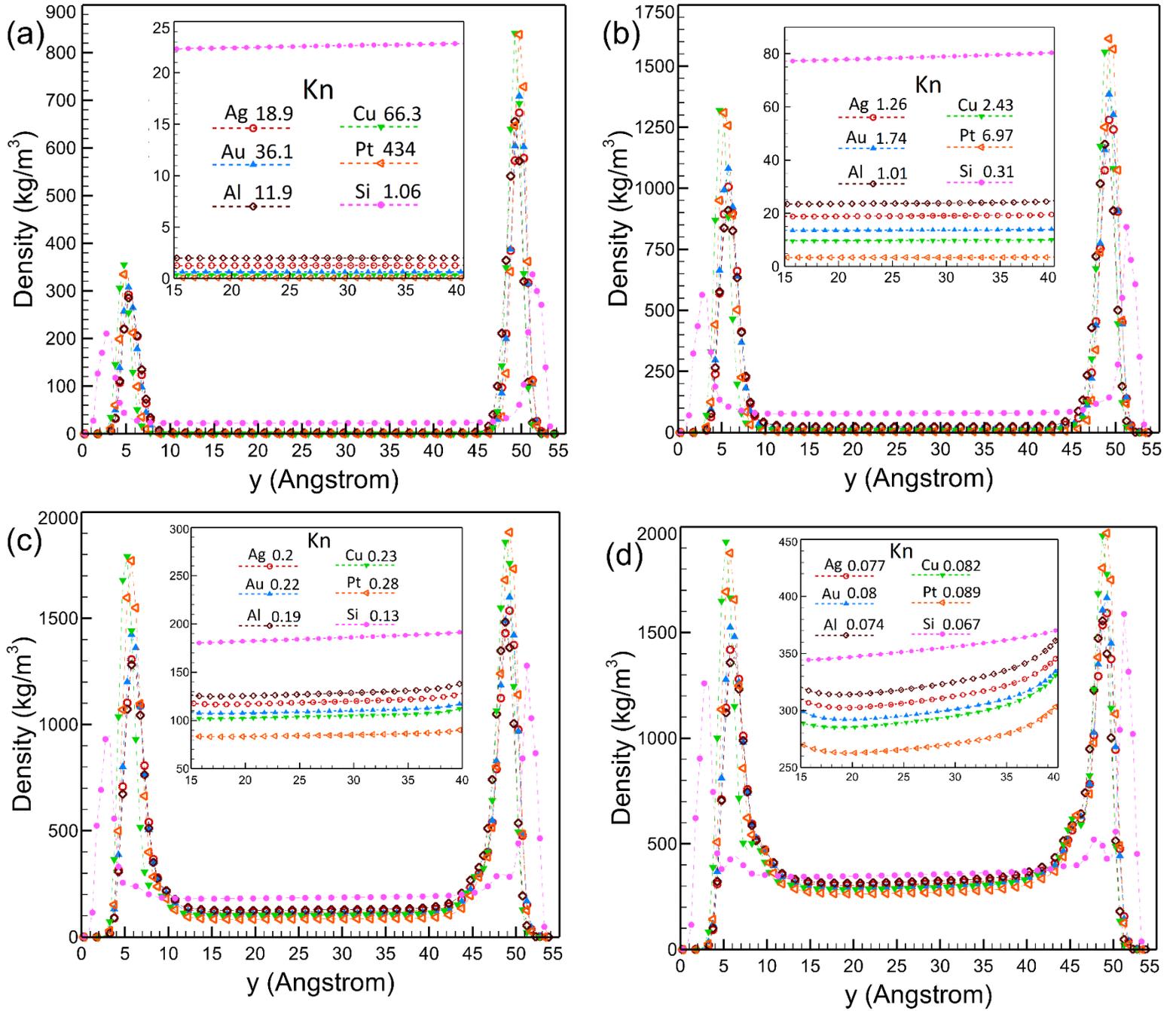

**Fig. 5.** Density distribution along the channel height for 10000 **(a)**, 30000 **(b)**, 60000 **(c),** and 100000 **(d)** gas atoms for constant gas atom number case (Knudsen numbers are computed based on the gas density in the middle of the channel)

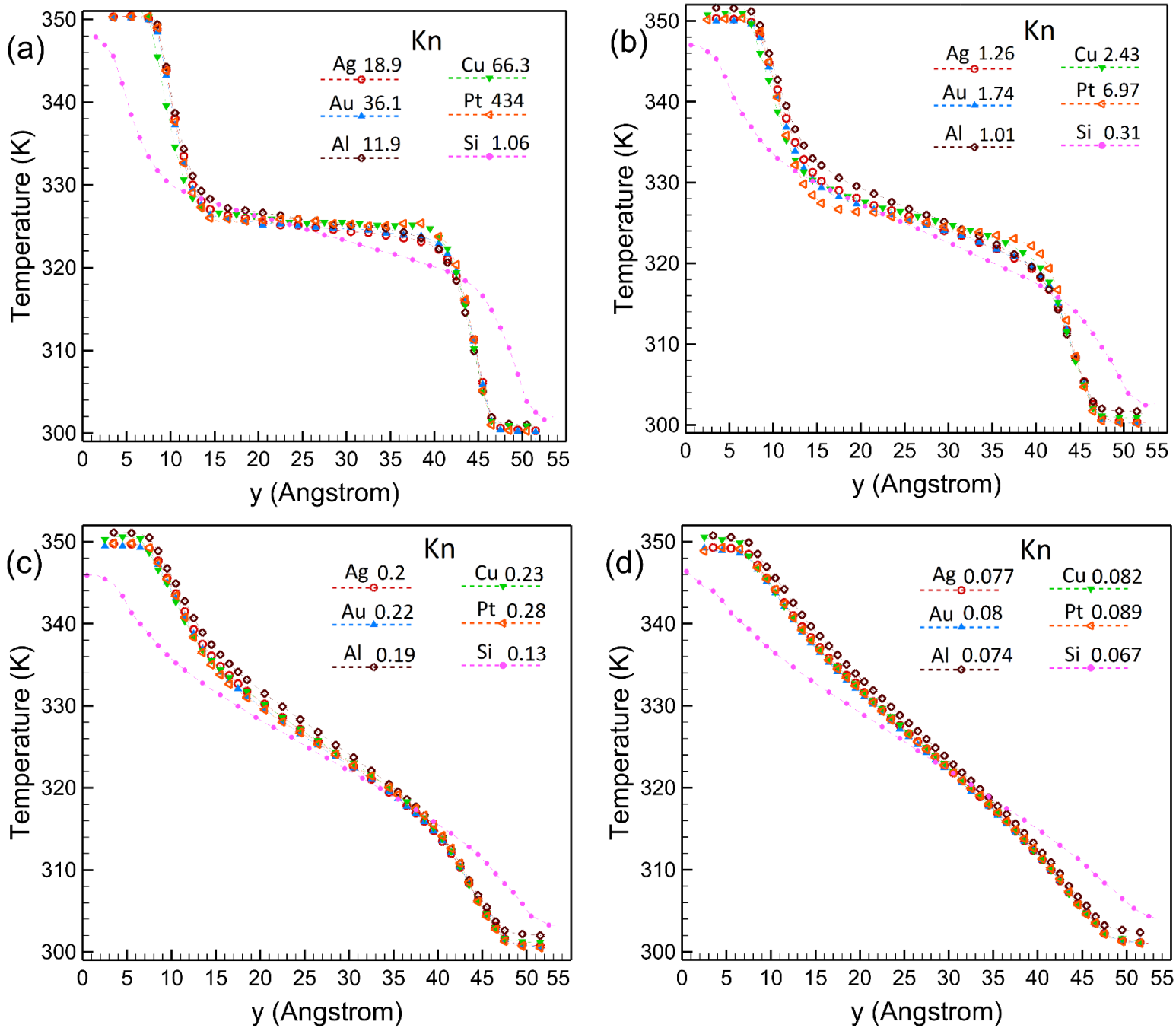

**Fig. 6.** Temperature distribution along the channel height for 10000 **(a)**, 30000 **(b)**, 60000 **(c),** and 100000 **(d)** gas atoms for constant gas atom number case (Knudsen numbers are computed based on the gas density in the middle of the channel)

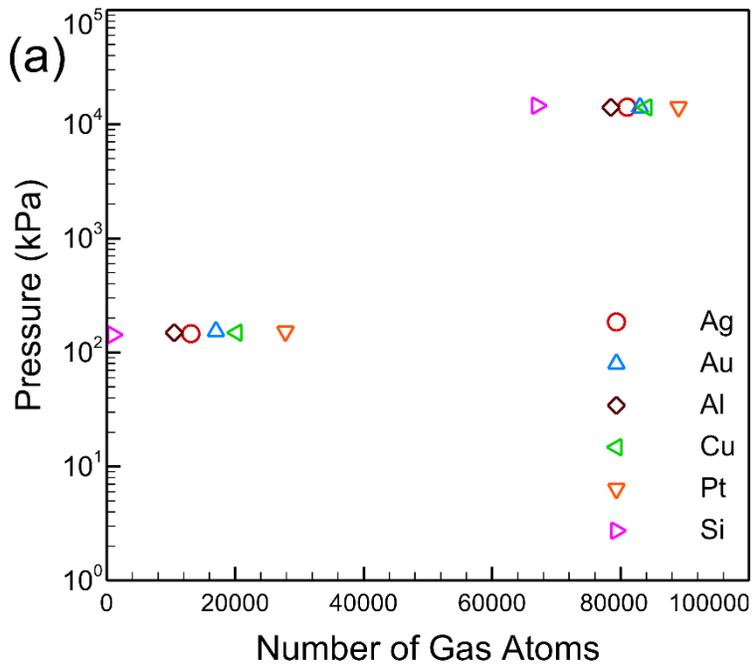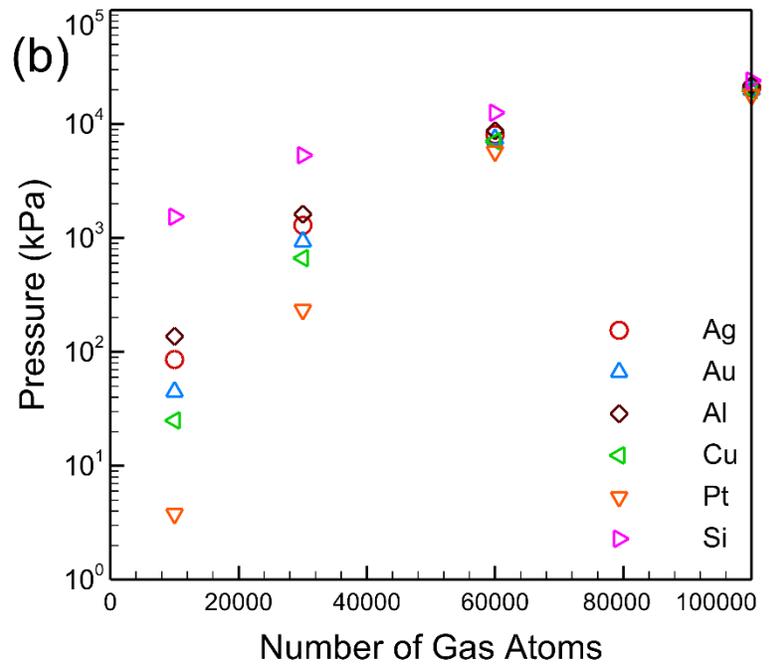

**Fig. 7.** Calculated pressure in the middle of the channel height for constant Knudsen **(a)** and constant number of gas atoms **(b)** cases

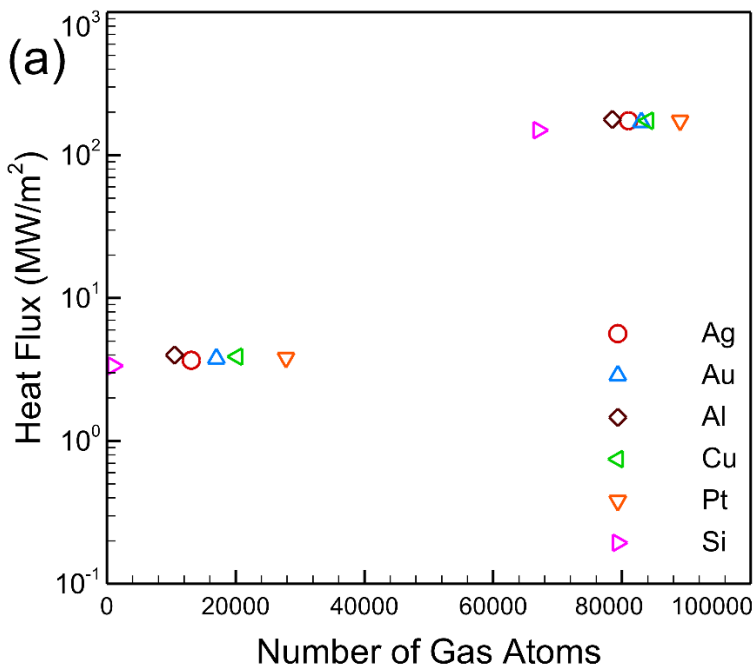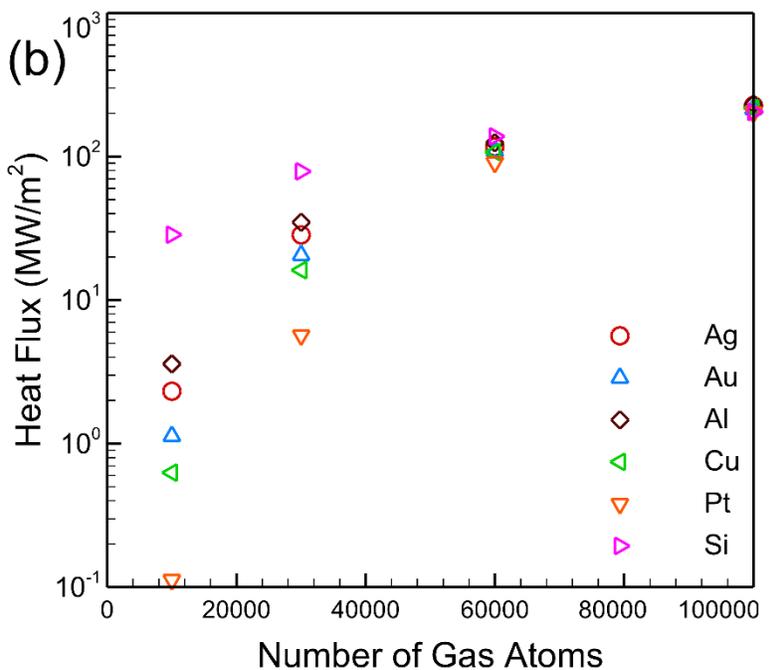

**Fig. 8.** Calculated heat flux in the middle of the channel height for constant Knudsen **(a)** and constant number of gas atoms **(b)** cases

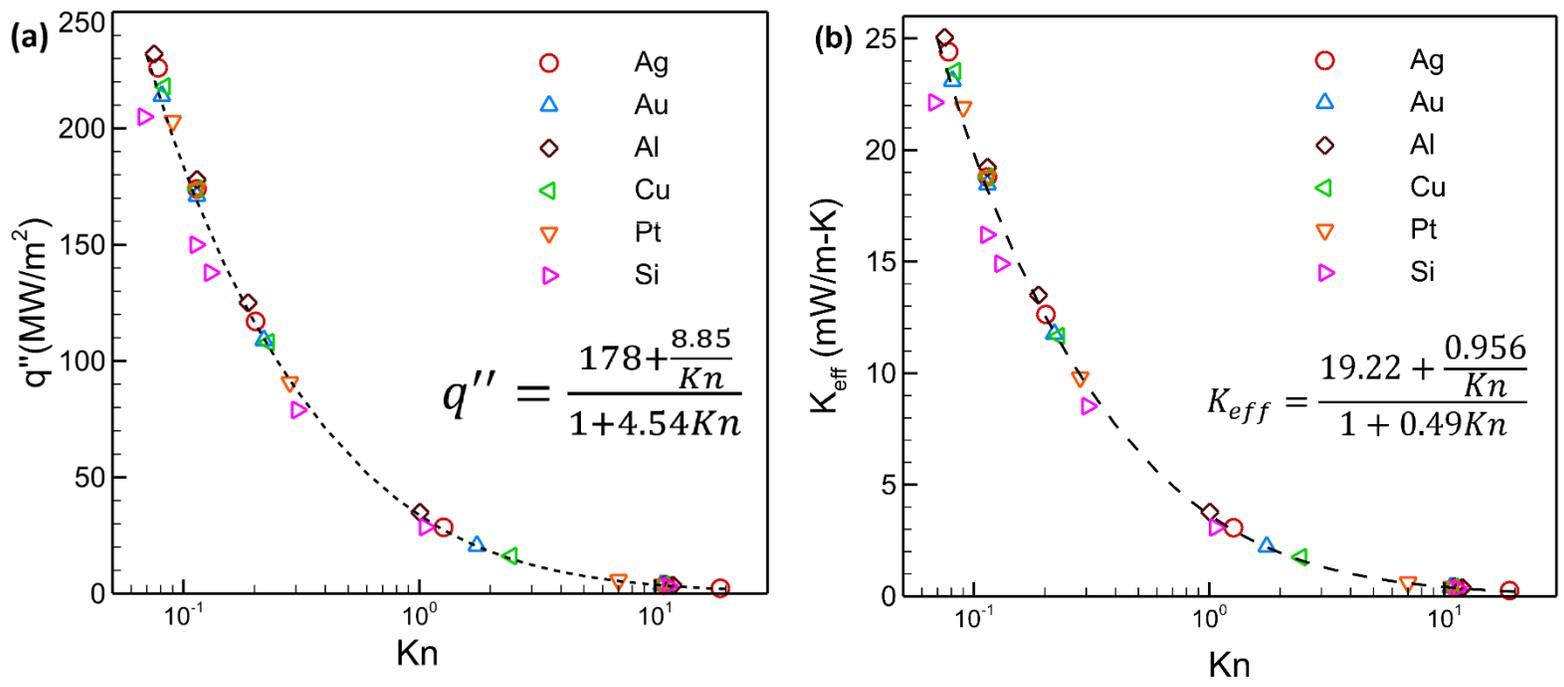

**Fig. 9.** Variation of induced (a) heat flux and (b) effective thermal conductivity as a function of Knudsen number for different wall materials and its corresponding analytical fitted curve

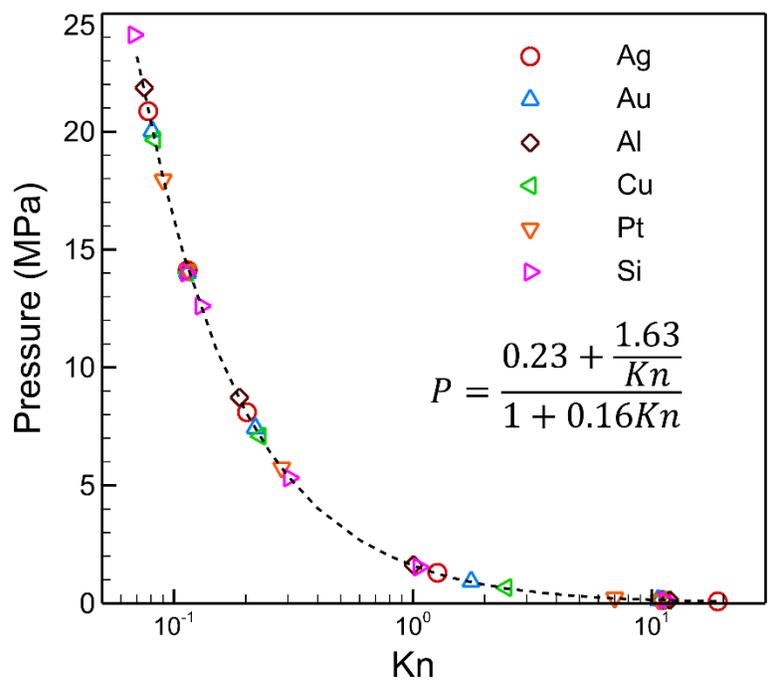

**Fig. 10.** Variation of pressure as a function of Knudsen number for different wall materials and its corresponding analytical fitted curve

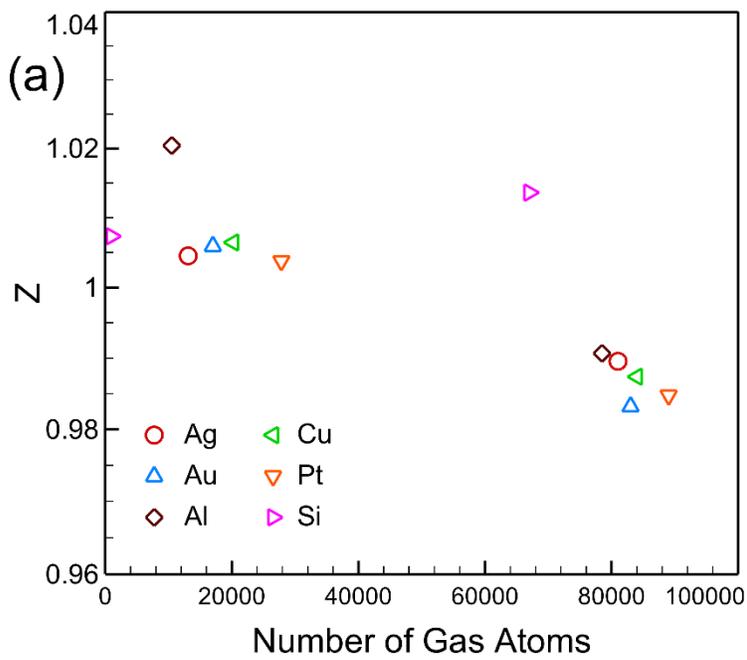 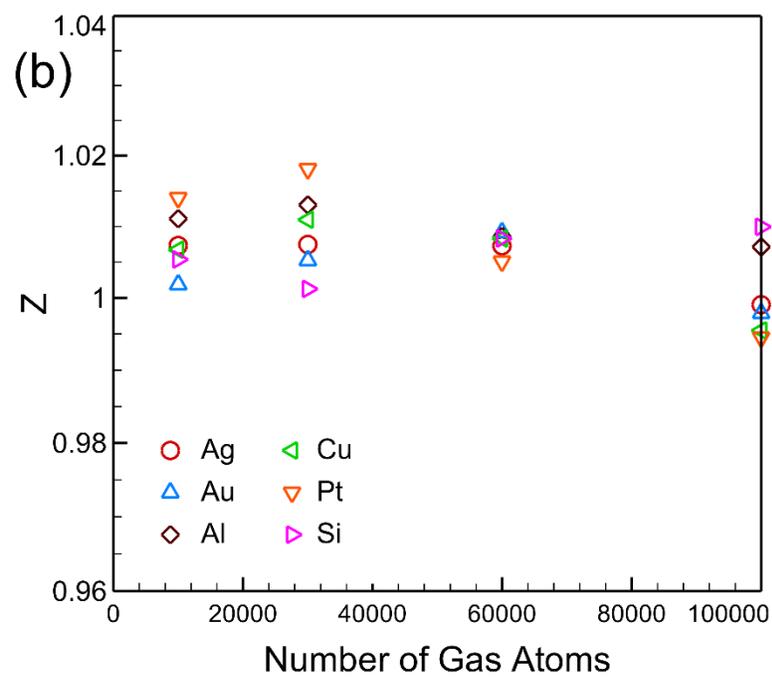

**Fig. 11.** Variation of compressibility factor (Z) in the bulk region of the gas for constant Knudsen (a) and constant number of gas atoms (b) cases

**Table 1** Interaction parameters in the simulation of the gas and wall atoms [66]

|  | $\sigma$ (nm) | $\varepsilon_{solid-gas}$ (J) | $\dfrac{\varepsilon_{solid-gas}}{\varepsilon_{gas-gas}}$ |
|---|---|---|---|
| $Ag$ [66] | 0.2955 | $7.25 \times 10^{-21}$ | 4.37 |
| $Au$ [66] | 0.295 | $7.8 \times 10^{-21}$ | 4.7 |
| $Al$ [66] | 0.293 | $6.8 \times 10^{-21}$ | 4.1 |
| $Cu$ [66] | 0.2616 | $7.37 \times 10^{-21}$ | 4.44 |
| $Pt$ [66] | 0.2845 | $9.5 \times 10^{-21}$ | 5.7 |
| $Si$ [67–69] | 0.338 | $2.56 \times 10^{-21}$ | 1.55 |

**Table 2** Physical and calculated properties of the argon gas and solid wall ($T_E = 0.806 T_D$)

|  | $T_E$ (K) | $K_S$ (N/m) | $m_{atom}$ (Kg) | $\dfrac{m_{wall}}{m_{gas}}$ | $\omega$ (THz) |
|---|---|---|---|---|---|
| $Ag$ [58,59] | 172 | 90 | $17.9 \times 10^{-26}$ | 2.7 | 2.24 |
| $Au$ [58,59] | 131 | 96 | $32.7 \times 10^{-26}$ | 4.93 | 1.71 |
| $Al$ [58,59] | 317 | 77 | $4.48 \times 10^{-26}$ | 0.68 | 4.15 |
| $Cu$ [58,59] | 248 | 110 | $10.5 \times 10^{-26}$ | 1.58 | 3.24 |
| $Pt$ [56,57] | 180 | 180 | $32.4 \times 10^{-26}$ | 4.89 | 2.36 |
| $Si$ [59] | 558 | 248 | $4.66 \times 10^{-26}$ | 0.703 | 7.29 |

**CONFLICT OF INTEREST**

There are no conflicts to declare.

**ACKNOWLEDGMENTS**

R.Rabani and A.Pishevar wish to thank the financial support of the Isfahan University of Technology.